%
\let\useblackboard=\iftrue
%
%
\newfam\black
\input harvmac.tex
\noblackbox
\def\Title#1#2{\rightline{#1}
\ifx\answ\bigans\nopagenumbers\pageno0\vskip1in%
\baselineskip 15pt plus 1pt minus 1pt
\else
\def\listrefs{\footatend\vskip 1in\immediate\closeout\rfile\writestoppt
\baselineskip=14pt\centerline{{\bf References}}\bigskip{\frenchspacing%
\parindent=20pt\escapechar=` \input
refs.tmp\vfill\eject}\nonfrenchspacing}
\pageno1\vskip.8in\fi \centerline{\titlefont #2}\vskip .5in}

\ifx\answ\bigans\def\tcbreak#1{}\else\def\tcbreak#1{\cr&{#1}}\fi
\useblackboard
\message{If you do not have msbm (blackboard bold) fonts,}
\message{change the option at the top of the tex file.}
\font\blackboard=msbm10 scaled \magstep1
\font\blackboards=msbm7
\font\blackboardss=msbm5
\textfont\black=\blackboard
\scriptfont\black=\blackboards
\scriptscriptfont\black=\blackboardss

\else

\fi
%
\def\yboxit#1#2{\vbox{\hrule height #1 \hbox{\vrule width #1
\vbox{#2}\vrule width #1 }\hrule height #1 }}
\def\fillbox#1{\hbox to #1{\vbox to #1{\vfil}\hfil}}
\def\ybox{{\lower 1.3pt \yboxit{0.4pt}{\fillbox{8pt}}\hskip-0.2pt}}

\def\comments#1{}

\def\p{\partial}

\def\half{{1\over 2}}
\def\Tr{{{\rm Tr\  }}}

\def\CF{{\cal F}}

\def\CN{{\cal N}}

\def\I{I}

\def\II{\relax{I\kern-.07em I}}

\def\inbar{\,\vrule height1.5ex width.4pt depth0pt}
\def\IZ{\relax\ifmmode\mathchoice
{\hbox{\cmss Z\kern-.4em Z}}{\hbox{\cmss Z\kern-.4em Z}}
{\lower.9pt\hbox{\cmsss Z\kern-.4em Z}}
{\lower1.2pt\hbox{\cmsss Z\kern-.4em Z}}\else{\cmss Z\kern-.4em
Z}\fi}
\def\IB{\relax{\rm I\kern-.18em B}}
\def\IC{{\relax\hbox{$\inbar\kern-.3em{\rm C}$}}}
\def\ID{\relax{\rm I\kern-.18em D}}
\def\IE{\relax{\rm I\kern-.18em E}}
\def\IF{\relax{\rm I\kern-.18em F}}
\def\IG{\relax\hbox{$\inbar\kern-.3em{\rm G}$}}
\def\IGa{\relax\hbox{${\rm I}\kern-.18em\Gamma$}}
\def\IH{\relax{\rm I\kern-.18em H}}
\def\II{\relax{\rm I\kern-.18em I}}
\def\IK{\relax{\rm I\kern-.18em K}}
\def\IP{\relax{\rm I\kern-.18em P}}

\def\p{\partial}

\font\cmss=cmss10 \font\cmsss=cmss10 at 7pt
\def\IR{\relax{\rm I\kern-.18em R}}

\def\sdtimes{\mathbin{\hbox{\hskip2pt\vrule
height 4.1pt depth -.3pt width .25pt\hskip-2pt$\times$}}}
\def\Tr{\rm Tr}

\def\BZ{\IZ}

\def\BC{\IC}

\Title{ \vbox{\baselineskip12pt\hbox{hep-th/9612062}
\hbox{CALT-68-2090}}}
{\vbox{
\centerline{Probing $F$-theory With Multiple Branes}}}
\centerline{Michael R. Douglas$^1$, David A. Lowe$^2$ and John H. Schwarz$^2$}
\medskip
\centerline{$^1$ Department of Physics and Astronomy}
\centerline{Rutgers University }
\centerline{Piscataway, NJ 08855-0849}
\centerline{\tt mrd@physics.rutgers.edu}
\medskip
\centerline{$^2$ California Institute of Technology}
\centerline{Pasadena, CA  91125, USA}
\centerline{\tt lowe, jhs@theory.caltech.edu}
\bigskip
\bigskip
We study multiple $3$-branes on an F theory orientifold. The
world-volume theory of the $3$-branes is $d=4$, $\CN=2$ $Sp(2k)$
gauge theory with an antisymmetric
tensor and four flavors of matter in the fundamental. The solution
of this gauge theory is found for vanishing bare mass of the
antisymmetric tensor matter, and massive fundamental matter.
The integrable system underlying this theory is constructed.

\Date{December 1996}
%
\lref\vafaf{C.~Vafa, ``Evidence for F-theory,'' Nucl. Phys. B469 (1996) 403;
 hep-th/9602022.}
\lref\question{P. Windey, private communication;
C. Vafa, private communication.}
\lref\sen{A.~Sen, ``F-theory and Orientifolds,'' Nucl. Phys. B475 (1996) 562;
 hep-th/9605150.}
\lref\bds{T. Banks, M. R. Douglas and N. Seiberg,
``Probing F-theory with Branes,'' Phys. Lett. B387 (1996) 278;
 hep-th/9605199.}
\lref\witsmi{E. Witten, ``Small Instantons in String Theory,'' Nucl. Phys.
B460 (1996) 541; hep-th/9511030.}
\lref\dansunone{U. H. Danielsson and B. Sundborg,
``Exceptional Equivalences in N=2 Supersymmetric Yang-Mills Theory,''
Phys. Lett. B370 (1996) 83;
hep-th/9511180.}
\lref\dansuntwo{U. H. Danielsson and P. Stjernberg,
``Notes on Equivalences and Higgs Branches in N=2 Supersymmetric
Yang-Mills Theory,'' Phys. Lett. B380 (1996) 68; hep-th/9603082.}
\lref\donwit{R. Donagi and E. Witten, ``Supersymmetric Yang--Mills
Theory and Integrable Systems,'' Nucl. Phys. B460 (1996) 299-334;
hep-th/9510101.}
\lref\sw{N.~Seiberg and E.~Witten, Nucl. Phys. B426 (1994) 19,
hep-th/9407087.}
\lref\swtwo{N.~Seiberg and E.~Witten, ``Monopoles, Duality and
Chiral Symmetry Breaking in N=2 Supersymmetric QCD,'' Nucl. Phys. B431 (1994)
484,
hep-th/9408099.}
\lref\markman{E. Markman, Comp. Math. 93 (1994) 255.}
\lref\hitchin{N. Hitchin, Duke Math. J. 54 (1987) 91.}
\lref\deboer{J. de Boer, K. Hori, H. Ooguri and Y. Oz,
``Mirror Symmetry in Three-Dimensional Gauge Theories, Quivers
and D-branes,'' hep-th/9611063.}
\lref\koblitz{N. Koblitz, ``Introduction to Elliptic Curves and Modular
Forms,'' Springer-Verlag New York, 1984.}
\lref\aha{O. Aharony, C. Sonnenschein, S. Yankielowicz, and S. Theisen;
``Field Theory Questions for String Theory Answers,''
hep-th/9611222.}
%
%
\newsec{Introduction}

F-theory may be defined as compactifications of Type IIB string
theory in which the dilaton and its Ramond-Ramond sector axion partner
are allowed to  vary over the internal space \vafaf.
F-theory compactification becomes
conventional perturbative string compactification for certain
 $\BZ_2$ orbifolds \sen.
Deformations away from the orbifold limit are described by the solution
of $d=4$, $\CN=2$ $SU(2)$ gauge theory with $N_f=4$ flavors of matter.
This can be understood physically by introducing a $3$-brane probe \bds:
it is the T-dual of the type \I\ $5$-brane with $SU(2)$ world-volume
gauge symmetry, and the matter comes from $3-7$ strings.
The variable dilaton-axion field $\tau$ is the low energy gauge coupling
on the $3$-brane.

It is natural to ask what happens when several parallel
$3$-branes are introduced \question.
As noted in \bds, in general adding $3$-branes
will change the background and thus the result need not have a probe
interpretation.  On the other hand, a system of multiple parallel $3$-branes
has a flat metric on moduli space, suggesting that the additional
$3$-branes might not affect the probe interpretation, and that
the matrix of coupling constants $\tau_{ij}$ might in fact be equal to
$\delta_{ij}\tau(z_i)$.

It turns out to be quite easy to show that this is true,
using an observation of Danielsson and Sundborg \dansunone.
The theory with $N_f$ $7$-branes and $k$ $3$-branes near an orientifold
point is the T-dual of the theory considered in
\refs{\witsmi}:
$Sp(2k)$ gauge theory with an antisymmetric tensor and $N_f$ flavors
of fundamental matter.
One can check that for $N_f=4$, this is a finite theory for any $k$,
so the basic physics is as in \sen\ even with multiple branes.

In this note we construct the solution of this theory. It turns
out to be convenient to describe this solution using the framework of
Donagi and Witten's solution of $SU(N)$ gauge
theory with massive adjoint matter \donwit.
We obtain in this way an integrable
system underlying the solution. An integrable system describing
$SU(2)$ gauge theory with four flavors of massive fundamental matter
is a special case of this construction.

\newsec{Solution of the Multiple 3-brane Probe Gauge Theory}

The first element of the solution of this theory is the
one-loop prepotential.
In terms of the roots $\alpha$ for the gauge group and weights $\lambda$
for the matter representations which appear, this is
\eqn\oneloop{
{\CF} \sim {i \over 4 \pi}
\sum _{\alpha} (\Psi \cdot \alpha )^2 \log{ (\Psi \cdot \alpha )^2  \over
\Lambda ^2} -
{i \over 4\pi}
\sum _{{\lambda},f} (\Psi \cdot {\lambda} -m_f)^2
\log{ (\Psi \cdot {\lambda} -m_f)^2
\over
\Lambda ^2} ~,
}
where $\Psi$ is the $\CN=2$ superfield describing the vector multiplet,
$\Lambda$ is the mass scale of the theory, and $m_f$ are bare mass terms.
Now, the key point is that for $Sp(2k)$,
the set of weights for the antisymmetric
tensor representation is a subset of the roots.  In a basis where
the weights of the fundamental are $\pm e_i$, the antisymmetric
tensor weights are $\pm e_i \pm e_j$, while the roots include all
of these and $\pm 2e_i$.  Thus the antisymmetric tensor contributions
simply cancel the off-diagonal roots.

The complete one-loop prepotential then is simply the sum of
those for independent $SU(2)$ factors with $N_f$ flavors,
\eqn\onelooptwo{\eqalign{
{\CF} &\sim {i \over 2 \pi}
\sum _{i} (2a_i)^2 \log{ (2a_i)^2  \over
\Lambda ^2} -
{i \over 4\pi}
\sum _{i,f} (a_i - m_{f})^2 \log{ (a_i - m_{f})^2  \over
\Lambda ^2}\cr & \qquad+  (a_i + m_{f})^2 \log{ (a_i + m_{f})^2  \over
\Lambda ^2}~.\cr}
}

Clearly the gauge coupling $\tau_{ij}=\p^2\CF/\p a_i\p a_j$
computed from this prepotential will be diagonal, and the $3$-branes
will behave as completely independent probes of the geometry,
to this approximation.

Danielsson and Sundborg's \dansunone\ observation is now that, given
a theory with a particular weak coupling behavior and corresponding
one-loop prepotential, the
Seiberg-Witten-type exact solution for the prepotential should be uniquely
determined.  If this is true, it means that the exact prepotential
for this theory must be equal to the sum of exact prepotentials
for $SU(2)$ with $N_f$ flavors.  Thus the $3$-branes
behave as independent probes of the exact geometry.

This can also be argued by considering the effects of a vev for the
antisymmetric tensor field, as in \dansuntwo.
This corresponds to separating the
$3$-branes in dimensions contained in the $7$-branes.
Clearly, for large separation the probes are independent.
Since hypermultiplet vevs cannot affect the vector effective Lagrangian
and $\tau$, this will remain true as we bring the probes together.

The answer to the original question turned out to be rather simple, and
we see that this particular matter content gives a nice generalization
of the finite $SU(2)$, $N_f=4$ gauge theory.

\newsec{A related non-D-brane gauge theory}

A more non-trivial generalization of the theory is
to allow a mass term for the antisymmetric tensor.
Although this has no direct brane interpretation, we are led by the
above to suspect
that the theory with this matter content might be tractable.
This theory has some similarity to $SU(N)$ with massive adjoint matter,
which was solved in \donwit. The techniques introduced
for that problem are useful here.
Donagi and Witten showed
that the solution of a $d=4$, $\CN=2$ gauge theory determines
a complex integrable system.  In favorable cases, this system can be
realized as a Hitchin system \hitchin\ as generalized by Markman \markman.
This construction automatically satisfies many of the physical
consistency conditions on the solution.
We expect the solution of the
theory with a massive antisymmetric tensor can be formulated in this
framework, though we will not obtain this solution in the present work.

As a first step toward this solution, we reformulate
the massless antisymmetric tensor theory in this language.
Given the solution of the $SU(2)$, $N_f=4$ theory \swtwo,
this solution follows easily, as might be expected from our
previous discussion.
For any scale-invariant $SU(2)$ theory, the
phase space for the integrable system is
the elliptic curve $E_\tau$ fibered over the moduli space $\IP^1$
of expectation values for $\bar u=\Tr \phi^2$.
This is equivalent to the $SU(2)$
Hitchin system on the curve $\sigma\cong E_\tau$.
Its phase space consists of gauge equivalence
classes of solutions of $F=\bar D\Phi =0$ with the symplectic structure
$\{A^a_{\bar z}(x),\Phi^b_z(y)\} = \delta^{ab}\delta(x-y)$.
The natural guess for the integrable system corresponding to
a theory with massive matter is a deformation of this.

The theory with massless adjoint matter had $SL(2,\BZ)$ symmetry and there is a
natural way to add a mass parameter preserving this symmetry: add a
`charged' source to the Hitchin equations at a single point
$\bar D\Phi = \mu\delta(x)$.
For $SU(2)$ there is a unique way to do this, while for $SU(N)$
with $N>2$ there is a unique way to do this which preserves the dimension
of the phase space.  This leads to the solution of \donwit.

The theory with $N_f=4$ massless fundamental flavors has
$Spin(8) \sdtimes SL(2,\BZ)$ global symmetry \swtwo, but this is broken to
$\Gamma(2)\subset SL(2,\BZ)$ for generic masses.
The full $Spin(8)\sdtimes  SL(2,\BZ) $ acts on the combined space of
moduli and mass
parameters, such that two $Spin(8)$ combinations
\eqn\invmass{
\eqalign{
R &=\half\sum_i m_i^2 \cr
N &= {3 \over 16} \sum_{i>j>k} m_i^2 m_j^2 m_k^2 -
{1\over 96} \sum_{i \neq j} m_i^2 m_j^4 + {1\over 96} \sum_i m_i^6 ~,\cr}
}
are invariant, while three other $Spin(8)$ invariants $T_i$ (only two of which
are independent)
\eqn\tmass{
\eqalign{
T_1 &= {1\over 12} \sum_{i>j} m_i^2 m_j^2 - {1\over 24} \sum_i m_i^4 \cr
T_2 &= -{1\over 2} \prod_i m_i - {1\over 24}  \sum_{i>j} m_i^2 m_j^2 +
{1\over 48} \sum_i m_i^4 \cr
T_3 &= -T_1-T_2 ~,\cr}
}
are permuted.

This suggests that we introduce charged sources proportional to the
conjugacy class $\mu = {\rm diag}(1,-1)$ in the Hitchin equations
at all four Weierstrass points
of the elliptic curve, which we can choose to be at positions
$\nu=0$, $1/2$, ${\tau/2}$ and $(1+\tau)/2$.
The coordinate $\nu$ is defined in
terms of the periodic
real parameters $\sigma_1$ and $\sigma_2$, each of period one, as
$\nu = \sigma_1+\tau \sigma_2$  and  transforms under
$SL(2,\BZ)$ as
$(\sigma_1, \sigma_2) \to (a \sigma_1 + b \sigma_2, c \sigma_1 + d \sigma_2)$,
for
\eqn\sltz{
\biggl(\matrix{a&b\cr
               c&d\cr} \biggr) \in SL(2,\BZ)~.
}
The source at the origin should be a singlet under triality of $Spin(8)$,
while sources at $1/2$, ${\tau/2}$ and $(1+\tau)/2$ should be permuted.
When the $T_i$ and $N$ vanish the system should reduce to that of the
massive adjoint case (\swtwo, 16.26).
This does not require all $m_i=0$ but rather the four masses can be
$(m,m,0,0)$.  The equivalence of the two theories can again be motivated
by comparing the one-loop prepotentials.
Define $L(a)=a^2 \log a^2$, then they are
\eqn\twocompare{\eqalign{
N_f=4~,\qquad \CF \sim & 2L(2a) - 4L(a) -2L(a-m) - 2L(a+m)  \cr
N=4~,\qquad \CF \sim &2L(2a') - L(2a'-m) - L(2a'+m)~.
}}
Then $L(2a)=4L(a)+{\rm regular}$, so
with $a=2a'$ these have the same singularities.
(This may be an interesting generalization of Danielsson and Sundborg's
observation.)
The sources at $1/2$, ${\tau}/2$ and $(1+\tau)/2$ should vanish in this
limit, in which case the integrable system reduces to that considered
by Donagi and Witten \donwit.

The subsequent analysis is facilitated by introducing the spectral curve
\eqn\specurv{
F(t,\tau)=\det (t-\Phi)=0~.
}
The sources in the Hitchin equations
translate into requiring specific poles for $F$, with residues
proportional to the charges. This, together with the requirements that
$F$ be of degree 2 (where we take $t$, $x$ and $y$ to be of degree
1,2 and 3 respectively) and that there be no additional singularities,
uniquely determines the function $F$ in terms of the four masses and
the parameter $\bar u$.

Let us use the Weierstrass representation for the curve $E_\tau$,
\eqn\weier{
y^2 = (x-e_1(\tau))(x-e_2(\tau))(x-e_3(\tau))~,
}
where the $e_i$ are roots of the polynomial $4 x^3 -g_2 x -g_3$,
which satisfy $\sum e_i=0$. Here $g_2 = 60 \pi^{-4} G_4(\tau)$ and
$g_3= 140 \pi^{-6} G_6(\tau)$ with $G_4$ and $G_6$ the
usual Eisenstein series \koblitz.
The three points $\nu=1/2$, ${\nu=\tau/2}$ and ${\nu=(1+\tau)/2}$
map into $(x=e_i,y=0)$, while $\nu=0$ maps into the point at infinity.
The spectral curve is then (generalizing \donwit, (3.7))
\eqn\suspecur{
F  = t^2 -\bar R x +\bar u - {P(x)\over { (x-e_1)(x-e_2)(x-e_3) }}~,
}
where $P(x)$ is a quartic polynomial in $x$ whose coefficients
are related to the five complex order parameters of the theory, which
we denote by barred variables
\eqn\polyn{
P= (\bar R x-\bar u)^3 \bigl( \bar T_1 (x-e_1) + \bar T_2 (x-e_2) +
\bar T_3 (x-e_3) \bigr)+
\bar N (\bar R x-\bar u)^4~.
}

The system of equations \specurv\ and \weier\ actually describes
a curve of genus two and thus the Jacobian is two dimensional.
However only a one-dimensional subspace of this is relevant for the
physics \donwit. As in the case considered there, the piece of the
Jacobian coming purely from $E_\tau$ must be projected out.
This is accomplished by modding out by the symmetry that
takes $t\to -t$ and $y \to -y$, which projects out the
form $dx/y$. The resulting curve is found by substituting the
invariant variables $z=t^2$ and $w=y t$ into \specurv\ and \weier. The
first equation allows us to eliminate $z$ while the second gives
\eqn\swcurv{
w^2 = (\bar R x-\bar u)(x-e_1)(x-e_2)(x-e_3) + P(x)~.
}

The form of this curve is rather different from
what was found in \swtwo. However, their equivalence is established
by the following
$SL(2,\BC)$ transformation of $x$
\eqn\sltwoc{
x = {{a x'+b}\over {c x'+d}}~,
}
with
\eqn\slparam{
\eqalign{
c&= {\bar R\over
\sqrt{(\bar u- \bar R e_1)(\bar u- \bar R e_2)(\bar u- \bar R e_3) }} \cr
a&= {\bar u c \over \bar R} \cr
b&= \bar R e_1 e_2 e_3 c \cr
d&={ (2{\bar u}^2 - \bar R^2 (e_1^2+e_2^2+e_3^2)) c \over 2\bar R}~.\cr}
}
This transformation also allows
us to match our five complex order parameters with those of \swtwo.
The relations are
\eqn\paramrel{
\eqalign{
\bar u &=u \cr
\bar R &= R \cr
\bar T_1 &= T_1 A (e_2-e_3)(u-R e_2)(u-R e_3)/B^2 \cr
\bar T_2 &= T_2 A (e_3-e_1)(u-R e_1)(u-R e_3)/B^2 \cr
\bar T_3 &= T_3 A  (e_1-e_2)(u-R e_1)(u-R e_2)/B^2 \cr
\bar N &= N  A^2/B^2~, \cr}
}
where we have defined $A=(e_1-e_2)(e_2-e_3)(e_3-e_1)$ and
$B=(u- R e_1)(u-R e_2)(u-R e_3)$.

This equivalence and \suspecur\ define the Hitchin
system which would be the definition of the integrable system
for $N_f=4$ in the Donagi-Witten framework.
The construction maintained manifest symmetry under $SL(2,\BZ)$ and triality,
but at a price: the sources in the Hitchin system are not linear in
the masses. This does not mean that the solution is inconsistent (after
all it is equivalent to that of \swtwo) but
rather that consistency is not manifest.
We have the freedom to add an exact form to the symplectic structure
and it might be possible to use this
to turn the description into another with linear sources, though probably
losing manifest $SL(2,\BZ)$.

The solution generalizes in an obvious way to the $Sp(2k)$ gauge theory
with a massless antisymmetric tensor and massive fundamental matter.
The charged sources are now taken to be proportional to
the conjugacy class $\mu= {\rm diag}(1,-1,\cdots,1,-1)$. Following
through the above construction, the spectral curve becomes
\eqn\spspec{
F= \prod_{i=1}^k F_2 (u_i)=0~,
}
where $F_2$ is given by \suspecur\ and the $u_i= \phi_i^2$. Here
we use the fact that the adjoint vev $\phi$ of $Sp(2k)$ can
be diagonalized as $\phi={\rm diag}(\phi_1,\cdots, \phi_k, -\phi_1,\cdots,
-\phi_k)$.  Note the physics is described by the part of the Jacobian
of the surface defined by \spspec\ and \weier\ which is invariant under Weyl
transformations. 

\newsec{Massive antisymmetric tensor matter}

We conclude with a few further comments on the generalization to massive
antisymmetric tensor matter.
There exists a value of the parameters which makes
the theory $SL(2,\BZ)$ invariant, and thus the theory should have the same
prepotential as $Sp(k)$ with a massive adjoint.
This is analogous to the equivalence of $SU(2)$ with four
flavors and masses $(m,m,0,0)$ and $SU(2)$ with a massive
adjoint.  Now the prepotentials of the two $Sp(k)$ theories are
\eqn\sppre{\eqalign{
N_f=4~,\qquad \CF=& \sum_{\alpha} L(\alpha\cdot a) - L(\alpha\cdot a-m_a)
+ \sum_{\tilde\alpha} L(2\tilde\alpha\cdot a) - 4L(\tilde\alpha\cdot a-m_f) \cr
N=4~,\qquad \CF=& \sum_{\alpha} L(\alpha\cdot a) - L(\alpha\cdot a-m)
+ \sum_{\tilde\alpha} L(2\tilde\alpha\cdot a) - L(2\tilde\alpha\cdot a-m)~,
}}
where $\alpha$ are the roots of the form $\pm e_i\pm e_j$,
and $\tilde\alpha$ are the weights $\pm e_i$. Here $m$, $m_f$ and $m_a$
are the masses of adjoint, fundmental and antisymmetric tensor matter,
respectively.
These prepotentials are the same (up to an irrelevant constant)
if $m_a=m$ and $m_f=m/2$.

We expect that the $Sp(2k)$ theory with massive adjoint can be expressed
as a Hitchin system with a single source. If this
is true then by including additional charged sources at all four
Weierstrass points it should be possible to construct the integrable
system describing the $Sp(2k)$ with a massive antisymmetric tensor
and massive fundamental matter.

We note that three-dimensional versions of these gauge
theories have been studied recently in \deboer. In that case, it is possible
to construct the moduli space for arbitrary mass parameters by using
the hyperk\"ahler properties of these spaces together with mirror
symmetry.

Multiple $3$-brane theories and their probe interpretation have also been
considered in \aha.

\medskip
\centerline{Acknowledgements}
M.R.D. and J.H.S. thank the Aspen Center for Physics, where this work was
begun.
D.A.L. wishes to thank R. Donagi and K. Landsteiner for helpful discussions.

This work was supported in part by DOE grants DE-FG02-96ER40559,
DE-FG03-92ER40701
and NSF grant PHY-9157016.

\bigskip

\listrefs
\end